\documentclass[sigconf,authorversion,nonacm]{acmart}

\usepackage{subfigure}
\usepackage{graphicx}

\AtBeginDocument{%
  \providecommand\BibTeX{{%
    \normalfont B\kern-0.5em{\scshape i\kern-0.25em b}\kern-0.8em\TeX}}}

\acmYear{2020}
\usepackage{xspace}

\makeatletter
\DeclareRobustCommand\onedot{\futurelet\@let@token\@onedot}
\def\@onedot{\ifx\@let@token.\else.\null\fi\xspace}

\def\eg{\emph{e.g}\onedot} 
\def\ie{\emph{i.e}\onedot} 
 
\def\etc{\emph{etc}\onedot}

\makeatother

\renewcommand\footnotetextcopyrightpermission[1]{} % removes footnote with conference information in first column
\begin{document}

\title{Towards an Interoperable Data Protocol Aimed at Linking the Fashion Industry with AI Companies}

%%%%  CHANGE IN TITLE NEEDED 
%%%% 
%%% 

% \title{Towards an Interoperable Data Exchange Protocol Aimed at Linking the Fashion Industry with AI Companies}

% Better Title add Exhange before 

\author{Mohammed Al-Rawi}
\email{alrawim@tcd.ie}
\affiliation{
  \institution{ADAPT Centre, Trinity College Dublin}
  \streetaddress{}
  \city{Dublin}
  \state{Ireland}
  \postcode{Dublin-2}
}
% \author{Dave Lewis}
% \email{dave.lewis@adaptcentre.ie}
% \affiliation{
%   \institution{ADAPT Centre, Trinity College Dublin}
%   \streetaddress{}
%   \state{Ireland}
%   \postcode{Dublin-2}
% }

\author{Joeran Beel}
\email{joeran.beel@tcd.ie}
\affiliation{
  \institution{ADAPT Centre, Trinity College Dublin}
  \streetaddress{}
  \city{Dublin}
  \state{Ireland}
  \postcode{Dublin-2}
}



\begin{abstract}
The fashion industry is looking forward to use artificial intelligence technologies to enhance their processes, services, and applications. Although the amount of fashion data currently in use is increasing, there is a large gap in data exchange between the fashion industry and the related AI companies, not to mention the different structure used for each fashion dataset. As a result, AI companies are relying on manually annotated fashion data to build different applications. Furthermore, as of this writing, the terminology, vocabulary and methods of data representation used to denote fashion items are still ambiguous and confusing. Hence, it is clear that the fashion industry and AI companies will benefit from a protocol that allows them to exchange and organise fashion information in a unified way. To achieve this goal we aim (1) to define a protocol called DDOIF that will allow interoperability of fashion data; (2) for DDOIF to contain diverse entities including extensive information on clothing and accessories attributes in the form of text and various media formats; and (3)To design and implement an API that includes, among other things, functions for importing and exporting a file built according to the DDOIF protocol that stores all information about a single item of clothing. To this end, we identified over 1000 class and subclass names used to name fashion items and use them to build the DDOIF dictionary. We make DDOIF publicly available to all interested users and developers and look forward to engaging more collaborators to improve and enrich it.

\end{abstract}


\begin{teaserfigure}
\centering
\includegraphics[width=\textwidth]{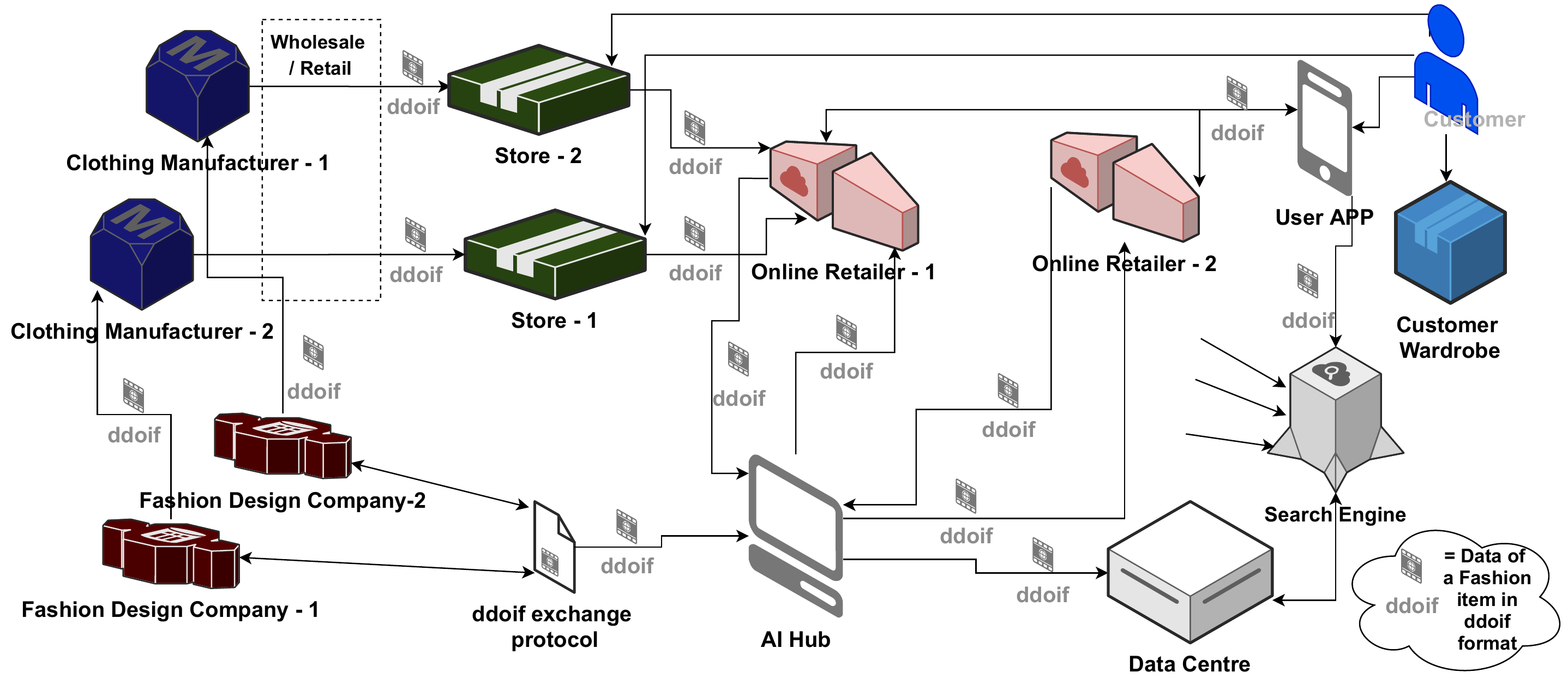}
  \caption{Business workflow that makes use of the Digital Data Exchange and Organisation in Fashion (DDOIF) protocol. There are many scenarios that DDOIF provides help with, for example; (1) the DDOIF file is generated at the clothing design stage and may be shipped with the item to the retailer and then to the customers app; (2) AI companies can either directly obtain clothing items in DDOIF format from manufacturers / retailers or use the DDOIF protocol to annotate their own data; (3) customers can use handset apps that host their electronic wardrobe items in DDOIF format, \eg, helping them picking the right clothing and style automatically; and (4) search engines can use the rich information that DDOIF files possess to provide accurate search results. See use-cases for more details.}
  \Description{Fashion industry.}
  \label{fig:teaser}
\end{teaserfigure}


%  Added be Rawi, to add page numbers
\settopmatter{printfolios=true}
% End of Rawi add

\maketitle

\begin{figure*}[htp]
  \centering
  \includegraphics[width=\linewidth]{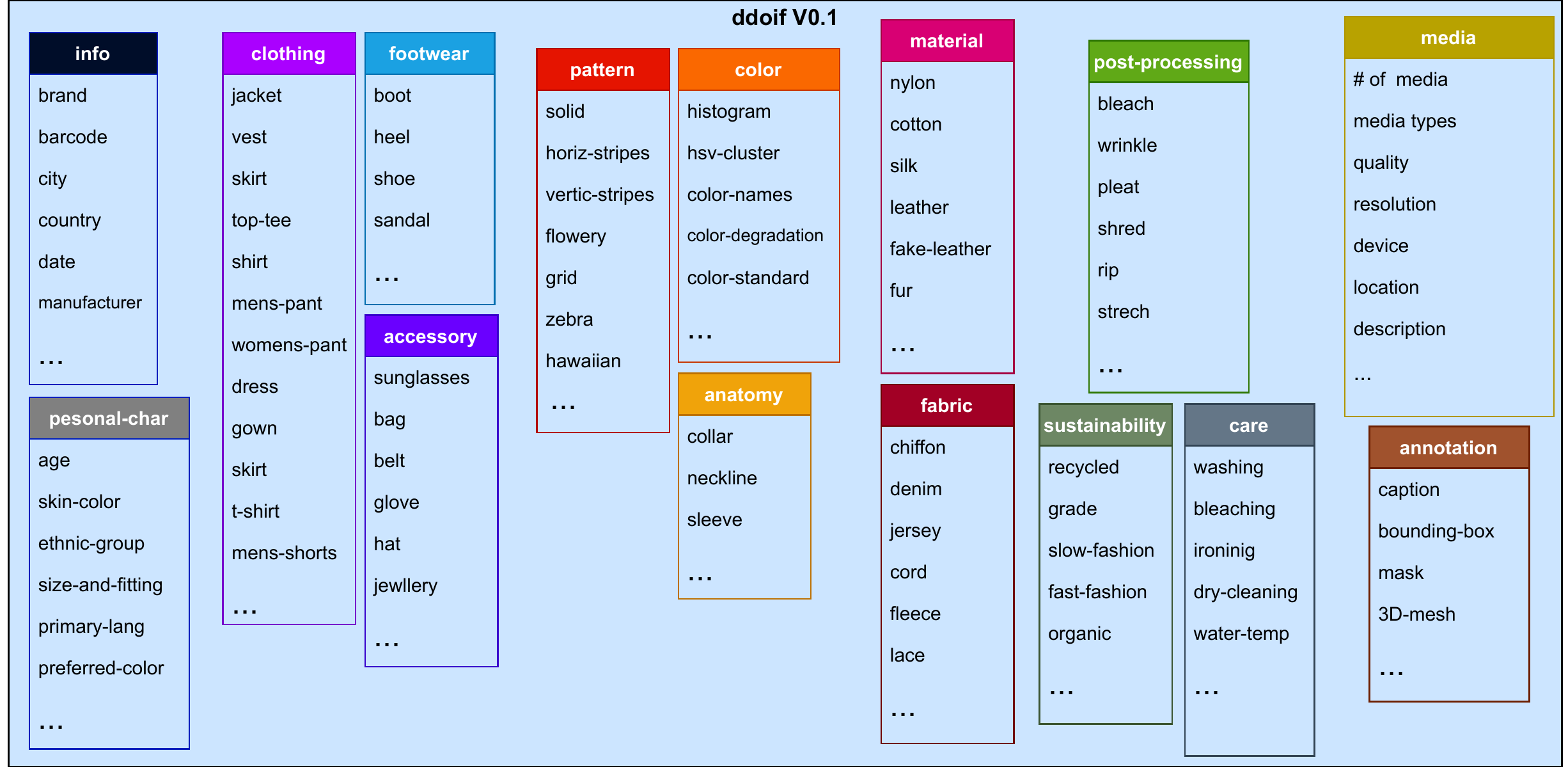}
  \caption{DDOIF classes. Due to lack of space, entries of classes are truncated. Best viewed in color, as colors mimic connections between classes shown in subsequent figures.}
  \Description{DDOIF Classes.}
  \label{fig:DDOIF}
\end{figure*}

\section{Introduction}
Nowadays, the fashion industry and related retail industries are racing to develop applications to serve them and their customers. This is usually done with the help of AI companies to provide different services; for example, recommender systems, AI based fashion design, applications enabling sustainability and slow fashion, trend forecasting, sizing and fitting, and brand protection \cite{Hou2019RecSys, GrokNet20, AmazonPatent}. In addition, organizations are heavily involved in research related to fashion aiming to develop different applications, \eg, to retrieve a particular fashion item based on a text query of clothing attributes and/or user preferences \cite{Ge2019DeepFashion2AV, Guo2019FashionIQ, Liang2016FashionCoParse, Zheng2018FashionModaNet, GrokNet20}. The impact of deep learning and the development of other AI methods provide companies with unprecedented means to achieve their goals. Amazon and StitchFix are providing their customers with the so called "Personal Styling" shopping, which is semi-assisted by fashion stylists \cite{AmazStch20, StichFix20}. Facebook is building a universal product understanding system where fashion is at its core \cite{FaceBk20, GrokNet20}. Zalando's researchers proposed in \cite{Studio2ShopFS} a model for finding pieces of clothing worn by a person in full-body or half-body images with neutral backgrounds. Some other companies are dedicated to fashion sizing and fitting \cite{size1, size2, size3}. However, it is unfortunate that these companies rely solely on AI to predict the features of fashion items that were already present during the design stage. These AI companies have to allocate non-trivial resources to forecast features that have been eliminated after design.

The fashion industry, which now enjoy enormous benefits from the information boom, still lack a proper data exchange format and relevant software that enable them exchange data interoperably. In fact, the fashion industry is currently using a diversity of in-house / proprietary data structures to identify and describe fashion articles and the vocabulary and terminology used to label their data are incongruous. While this is understandable due to the descriptive complexity of the fashion items, this diversity hinders cooperation within and between companies, complicates and constrains data interoperability and the use of artificial intelligence (AI) in fashion systems. AI companies involved in fashion have identified this data limitation problem without a clear view of how to solve it \cite{HeuriTech}.

In this work we aim to (1) design a Digital Data Exchange and Organisation in Fashion (DDOIF) protocol; (2) collect and organise the fashion vocabulary and terminology into DDOIF dictionary; (3) create relevant API (DDOIF API) that enables handling, importing and exporting the fashion data according to DDOIF; and (4) enable the API to encapsulate fashion data of a single item into a single file. A generic business workflow making use of DDOIF is illustrated in Figure~\ref{fig:teaser}, where DDOIF files can be shipped to different stakeholders. 

It is clear that the information one piece of clothing might contain is immense. The main classes that DDOIF may contain are depicted in Figure ~\ref{fig:DDOIF}. It may therefore be impractical to use crowdsourcing to non-uniformly, and possibly inaccurately, annotate items that have already been fabricated according to very specific details and attributes. This is because all features and annotations are available at the design stage. Shipping these attributes in DDOIF format with the item will save a lot of time and effort, and will provide more accurate item description.  Moreover, a protocol that can be used to build AI based applications should contain vital information about each item; \eg, vocabulary and terminology, media, visual attributes, design patterns and colours, material, type and texture of the fabric, colour and style matching criteria, recommendations according to fashion stylists, customer preferences, description, outfit type (\eg, outdoor, casual, nightwear), \etc. The existence of such information will help the industry build much better services that are accurate and robust.

\begin{figure*}[htp]
  \centering
  \includegraphics[width=0.85\linewidth]{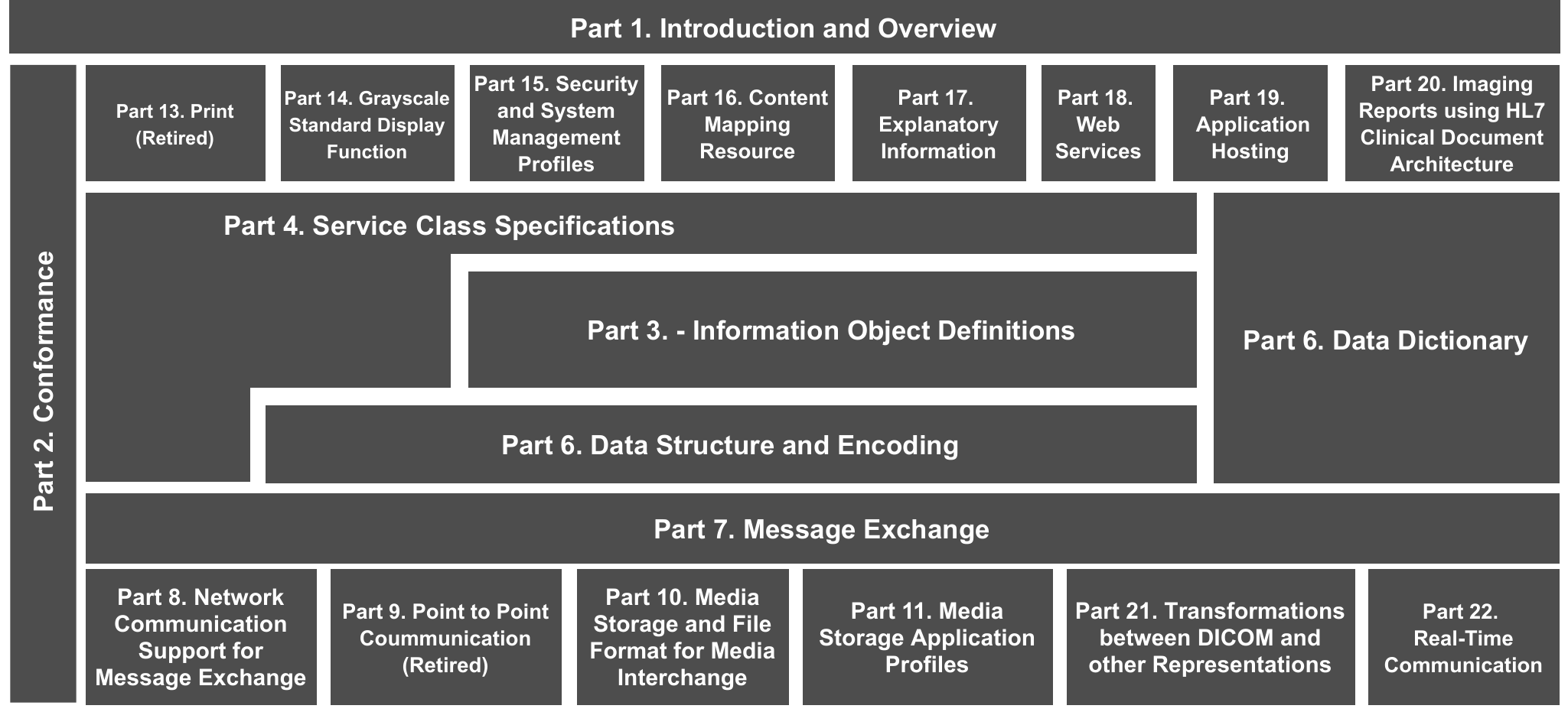}
  \caption{DICOM Structure.}
  \Description{DICOM.}
  \label{fig:DICOM}
\end{figure*}

\section{Literature Review}

\noindent \textit{\textbf{1. The way AI companies build applications.}}
AI companies build applications via a data-driven approach. Their main first step is to build a fashion dataset by annotating the fashion images ~\cite{Ge2019DeepFashion2AV, Guo2019FashionIQ, Liang2016FashionCoParse, Zheng2018FashionModaNet}. The annotation is usually done manually to segment the images and / or define their main attributes using some crowdsourcing platform, such as the Amazon Mechanical Turk ~\cite{Crowston2012AmazonMT}. Although these annotated datasets provide good benchmarks for implementing AI fashion research and other related applications, the annotation is restricted to a very limited clothing categories, costly, and prone to human error. Moreover, the naming and categories of fashion items across different datasets are not uniform and may vary. Nonetheless, the exact and accurate naming and clothing attributes were already available at the time of design and/or manufacturing. Unfortunately. the information that the manufacturer sends to retailers is not only limited, but also not uniform across different manufacturers because it does not follow a unified protocol. It would be highly beneficial to standardise this information and ship it along the clothing item to the retailers, and customers' apps as well.

\noindent \textit{\textbf{2. The current situation of fashion standards.}}
ICT standards are clearly very useful in many areas, and the fashion community can benefit in a similar way. Among the many useful ICT standards there is the JPEG standard (\cite{Penn92}; Rec. ITU T.81 | ISO/IEC 10918), which is one of the most successful multimedia standards to date. The JPEG standard (the Joint Photographic Experts Group) specifies the basic encoding technology and includes several options for encoding images, compliance testing, defining a set of extensions to the coding technologies, specifying the compression types, file exchange format, registration authorities, \etc. Another relevant ICT standard is the DICOM standard \cite{Dicom, Kahn2007DICOMAR} (Digital Imaging and Communications in Medicine). DICOM aims to store, retrieve, print, process, and display medical imaging information; thus, making medical imaging information interoperable. Another ICT standard is DICONDE (Standard Practice for Digital Imaging and Communication in Nondestructive Evaluation) of the American Society for Testing and Materials \cite{AstmDICONDE}. To provide an example of the complexity of ICT standards, the DICOM standard contains more than 4,000 tags and corresponding values. Figure~\ref{fig:DICOM} illustrates the structure of DICOM and the relationship between its parts.

To our knowledge, there is no ICT standard or protocol that comprehensively deals with fashion terminology, vocabulary, and media formats. There are, however, a few ISO standards for the garment industry, to name a few: (1) ISO/TC 133 \textit{Clothing Sizing Systems - Size Designation, Size Measurement Methods and Digital Fittings} \cite{isoDigiFitt}; and (2) ISO 18163:2016 \textit{Clothing — Digital fittings — Vocabulary and Terminology used for the Virtual Garment} \cite{iso2016}. ISO/TC 133 is a standardization of a system of size designations resulting from the establishment of one or more sizing systems for clothes. It is based on size designation, body size measurement methods for clothing and for digital garment fitting. ISO 18163:2016 specifies the data attributes and formats required for the creation of virtual garments, facilitating clear and synchronized communication of terminology. The major advantage of this standard is that it supports online consumers, fashion designers, manufacturers and retailers who have an interest in the style and fit of apparel. While ISO/TC 133 is only dedicated to sizing and fitting, the shortages of ISO 18163:2016 and ISO/TC 133, which may be closest to the protocol that we aim to propose lacks, among other things, detailed subclass names and anatomy of garments. Both standards lack associating the vocabulary and attributes to media formats; and importantly, how these items can be connected to the emerging AI applications. Moreover, these ISO are just guiding standards and not like the DICOM or DICONDE that provide comprehensive solutions to the related industries. Having some ICT protocols or standards for fashion analogous to DICOM or DICONDE will result in a very big impact on the fashion industry.

\begin{figure}[!htb]
 \centering
     \includegraphics[width=0.7\linewidth]{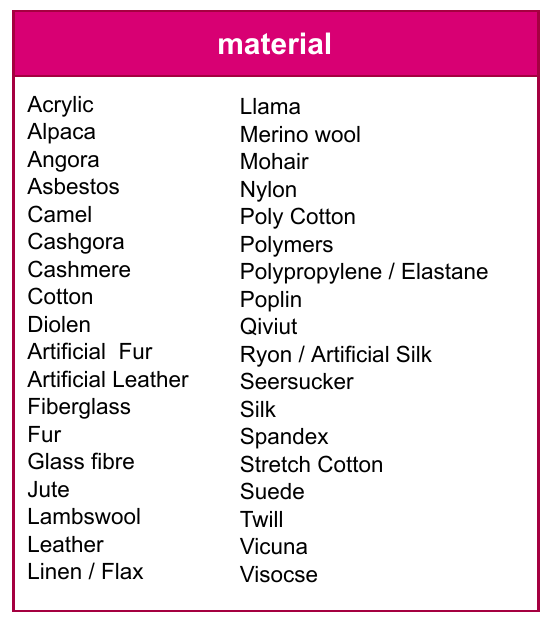}
      \caption{Material class.} \label{Fig:material-class}
\end{figure}

\begin{figure}[!htb]
   \begin{minipage}{0.6\linewidth}
     \centering
     \includegraphics[width=0.7\linewidth]{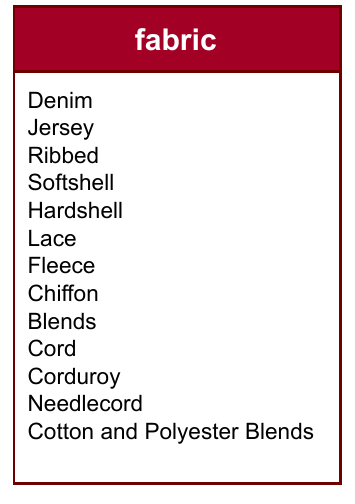}
     \caption{Fabric class. 85 classes list truncated due to lack of space.} 
     \label{Fig:fabric-class}
   \end{minipage}\hfill
   \begin{minipage}{0.35\linewidth}
     \centering
     \includegraphics[width=1.1\linewidth]{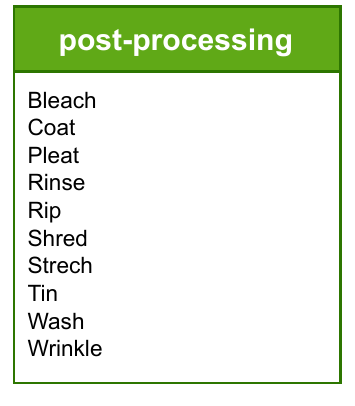}
     \caption{~Post-Processing class.} 
     \label{Fig:Post-Processing-class}
   \end{minipage}
\end{figure}

\begin{figure*}[htp]
  \centering
  % width=\linewidth,height=0.9\textheight
  \includegraphics[scale=0.73]{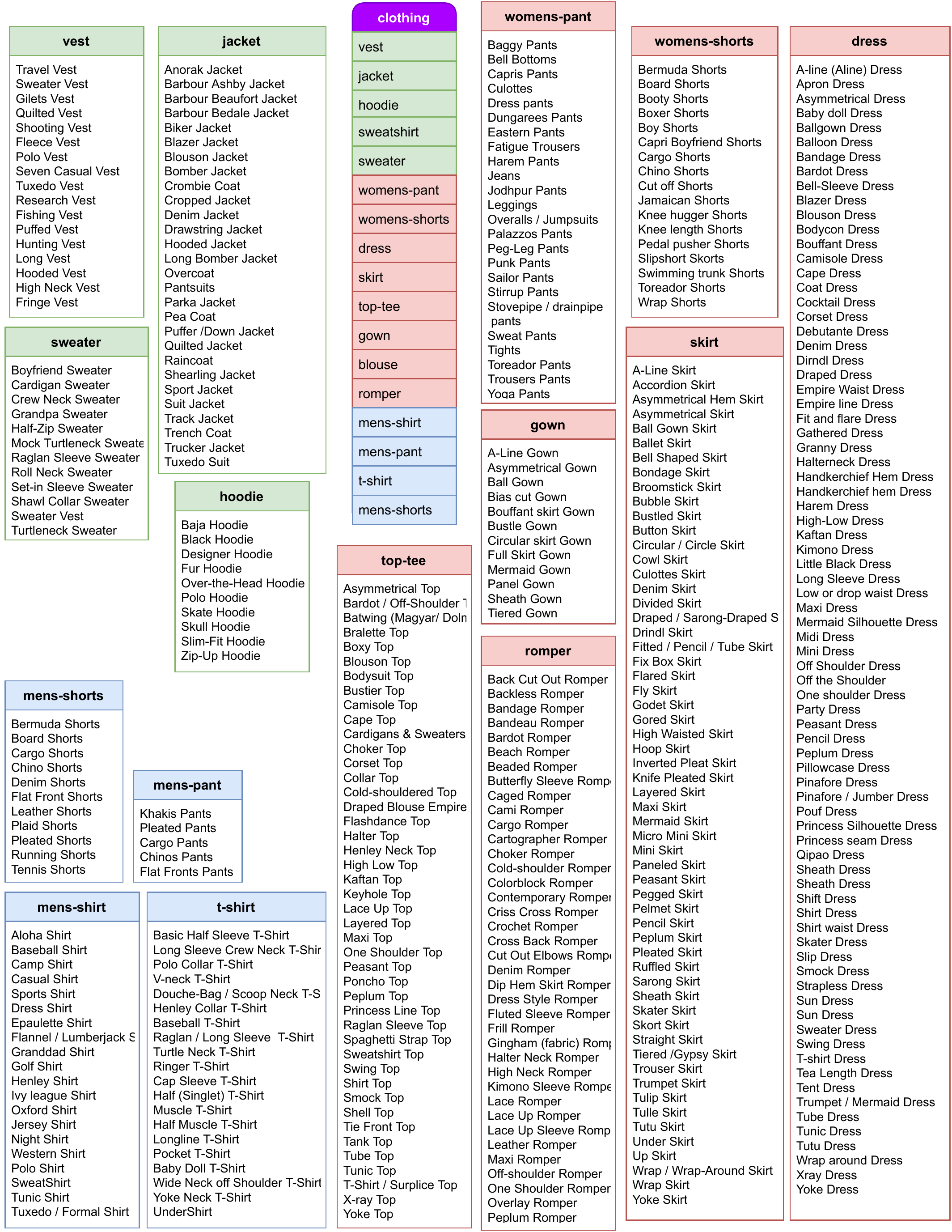}
  \caption{Clothing Classes. Romper items are truncated due to lack of space. Best viewed in color, as colors mimic connections between classes. The full list can be accessed in \cite{DDOIFV0.1}.}
  \Description{Clothing classes.}
  \label{fig:clothing_classes}
\end{figure*}

\section{The DDOIF Protocol}

DDOIF facilitates the interchange of information between applications in the fashion industry environments by specifying:

\begin{itemize}
\item{\verb|Naming|}: The vocabulary and terminology of fashion articles, and relevant AI annotations.
\item{\verb|Format|}: Fashion items characteristics and associated media formats.
\end{itemize}

We therefore conceptualize DDOIF by defining its terminology and vocabulary structure in one part (DDOIF Dictionary) and a file structure (DDOIF File) that encapsulates a fashion item based on the dictionary part. While the first part can be used by a user as a reference to ensure fashion items are correctly named and structured, the second part can be used to exchange information between different parties.

\subsection{DDOIF Dictionary}
\begin{figure}[!htb]
  \centering
  \includegraphics[width=0.8\linewidth]{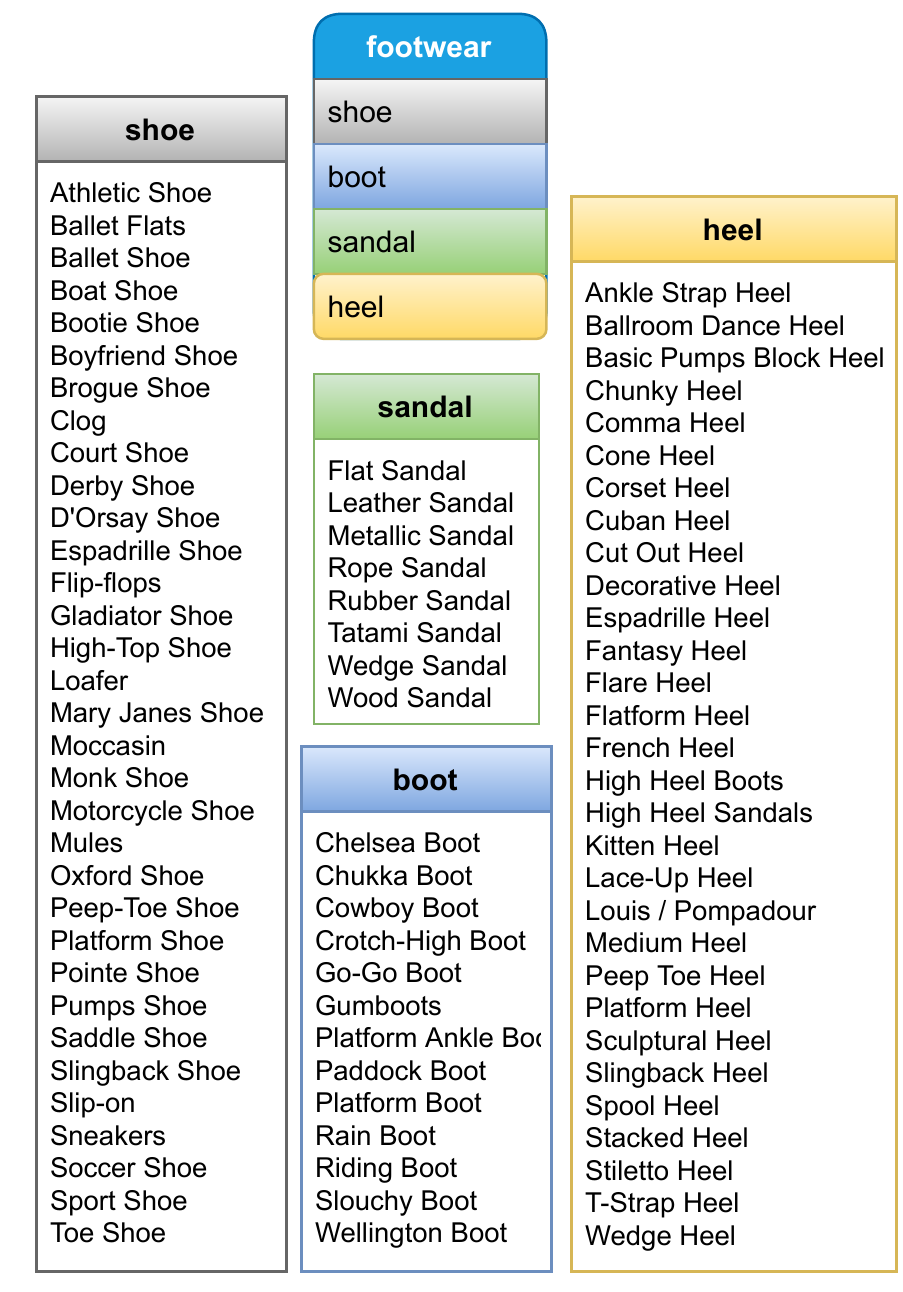}
  \caption{Footwear Classes. Best viewed in color, as colors mimic connections between classes.}
  \Description{Footwear Class.}
  \label{fig:Footwear_class}
\end{figure}

\begin{figure}[!htb]
  \centering
  \includegraphics[width=1\linewidth]{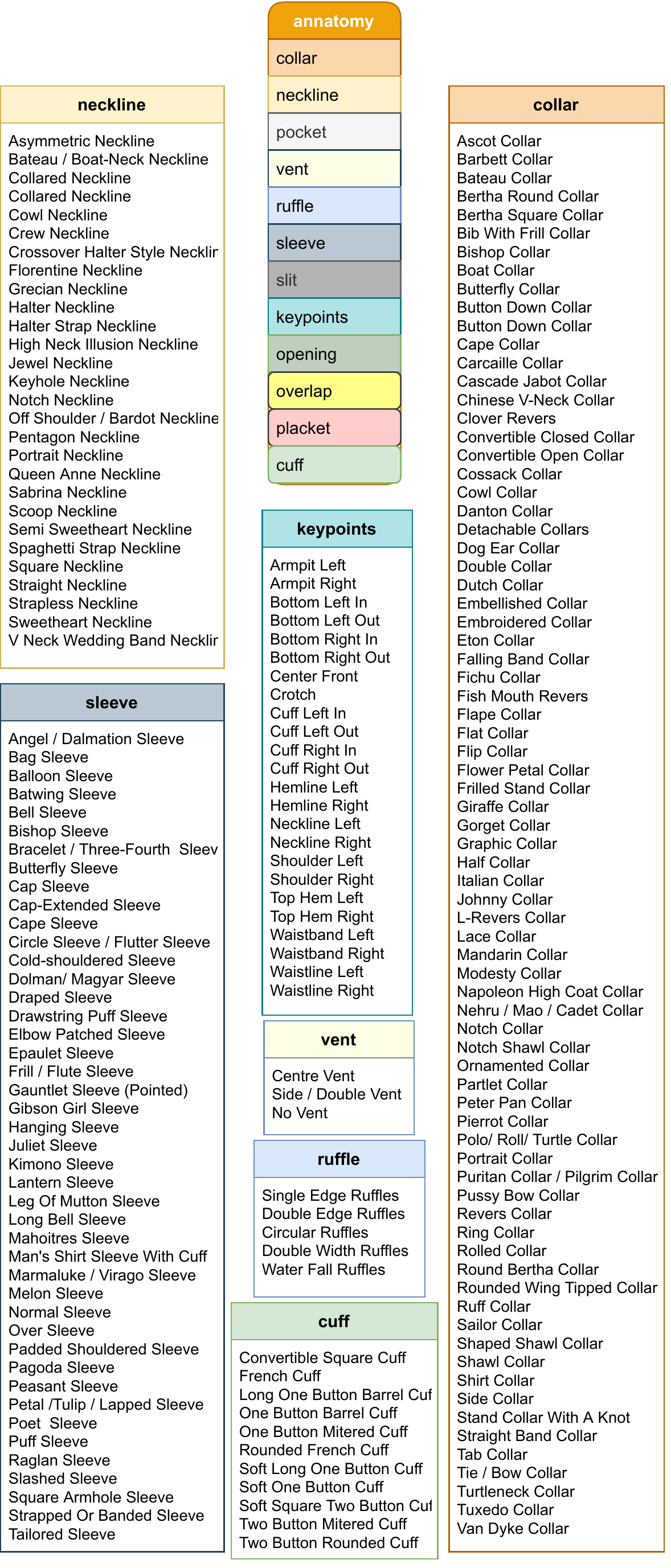}
  \caption{Anatomy Classes. Best viewed in color, as colors mimic connections between classes.}
  \Description{Anatomy Class.}
  \label{fig:Anatomy_class}
\end{figure}

\begin{figure}[ht]
  \centering
  \includegraphics[width=0.8\linewidth]{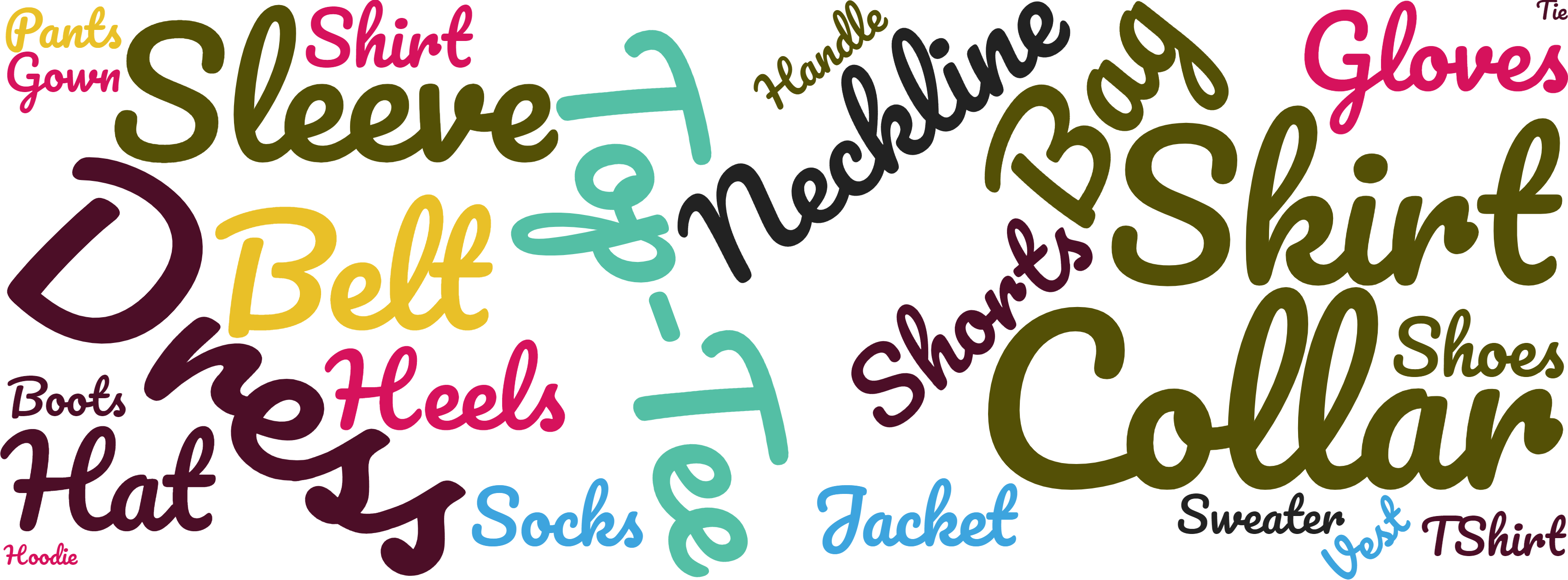}
  \caption{DDOIF Clothing, Accessory, and Anatomy Name Cloud Illustration. We only show the highest frequencies. Colors have no meaning here other than providing a good visualisation.} 
  \Description{DDOIF Cloud.}
  \label{fig:DDOIF-cloud}
\end{figure}

We defined the fashion item classes into levels; the first-level is the main class, the second-level is a subclass from the first-level, and so on. For illustration, first-level classes are shown in Figure ~\ref{fig:DDOIF} (the colored field names), second-level classes (\ie subclasses) are also listed under each first-level class. Then in Figure ~\ref{fig:clothing_classes}, we demonstrate the clothing class down to the third-level. As an example, the clothing class (first-level) has subclasses Dress and Skirt (second-level). Then, Dress class has third-level classes according to the type name of dresses, \eg A-line Dress and Apron Dress, \etc. If the A-line dress class has some distinctive features, it will generate other subclasses. Hence, each level $m$ entry in the DDOIF dictionary has the following building block:

 \begin{center}
 \hspace{-6cm} $class-level_m$ \\
 \hspace{-5cm} $class_1-level_{m-1}$ \\
 \hspace{-5cm} $class_2-level_{m-1}$ \\
 \hspace{-5cm} $...$  \\
 \hspace{-5cm} $class_i-level_{m-1}$ \\
 \hspace{-5cm} $...$ \\
 \hspace{-5cm} $class_n-level_{m-1}$ \\
 \hspace{-5.5cm} $description_m$ 
\end{center}

\noindent where $class-level_m$, which is the parent class, contains all the possible child / sub classes in addition to 'description' that may contain some information about  $class-level_m$. Each (child / sub) class is then entitled to act as a parent class, and so on. We collected several fashion class and subclass names from many Internet sources, \eg, fashion style and online retail sites \cite{GoogleArt, Zalando, PopOptiq, AmazonFashion}. We also used ISO 18163:2016 \cite{iso2016} to ensure that the main classes comply with ISO, if available and/or needed. It must be noted that ISO 18163:2016 only and partially supports first-level classes. We designate all first-level classes in lowercase letters, otherwise, the first letters of the subclasses are capitalized. We then built the DDOIF dictionary according to the structure described above using YAML, JSON, and XML. Future releases will contain additional entries by dynamically adding more classes as they appear, depending on fashion evolution. We present in Figures~\ref{fig:DDOIF},~\ref{Fig:material-class},~\ref{Fig:fabric-class},~\ref{Fig:Post-Processing-class},~\ref{fig:clothing_classes},~\ref{fig:Footwear_class}, and~ \ref{fig:Anatomy_class} some DDOIF classes. All classes can be accessed in \cite{DDOIFV0.1}. These classes give an idea about the terminology and vocabulary that DDOIF uses. In all diagrams, we use colors to mimic connections between main and sub classes. Interestingly, the number of classes of women's clothing items are much higher than those of men, see Figure \ref{fig:clothing_classes} pink versus light blue classes. This accords with women's interest and passion for fashion. This is also clear in Figure~\ref{fig:DDOIF-cloud} where the frequency distribution of classes is well presented.

\subsection{DDOIF API}
The API is a software used to enable importing and exporting fashion items according to the DDOIF protocol. Consequently, we designed the DDOIF file as follows:

\subsubsection{DDOIF File Header}
We designed the DDOIF file to start with a 10-byte signature (\ie Magic Number or Magic Bytes) containing the following:

\begin{enumerate}
\item \texttt{0x89} has the high bit set to detect systems that do not support one byte transmission data.
\item \texttt{0x44 '0x44' '0x4f' '0x49' '0x46'}	in ASCII, the letters DDOIF enable identifying the format easily if it is viewed in a text editor.
\item \texttt{0x0D 0x0A} a DOS-style line ending (CRLF) to detect DOS-Unix line ending conversion of the data.
\item \texttt{0x1A}	A byte that stops display of the file under DOS when the command type has been used—the end-of-file character.
\item \texttt{0x0A}	A Unix-style line ending (LF) to detect Unix-DOS line ending conversion.
\end{enumerate}

\subsubsection{Reserved bytes} 16 bytes are reserved for future additions, in case needed. Currently filled with 0 values.

\subsubsection{Data Chunks}
We designed DDOIF file in a manner that encapsulates textual data denoting fashion attributes, media types of a single piece of clothing and the other information according to the DDOIF dictionary. The textual data chunc consists of:

\begin{enumerate}
\item 4 bytes store the length in bytes (let the length be denoted by \textit{N}) of the DDOIF textual description according to the DDOIF dictionary.
\item \textit{N} bytes store the classes and attributes of the item according to the DDOIF dictionary.
\end{enumerate}

\noindent 
It is well known that the design of fashion items may articulate different types of media data; for example, 3D objects, 2D images at different views (front, back, side, \etc) and even videos. Each media format will constitute a data chunk in the DDOIF file. Consequently, we store the media data chunks sequentially, and each will contain the following stream of bytes:

\begin{enumerate}
\item 8 bytes store the format name (i.e type) of the media, \eg, JPG / JPEG , PNG, TIFF, STL, OBJ, 3DS, \etc. This string will be needed to know the type of stored media and to decode it once it is read. We choose 8 bytes instead of 4 to enable future extension to formats that may have more than 4 letters.
\item 4 bytes to store the length of the Media Buffer, let it be denoted as \textit{M}.
\item \textit{M} bytes store the media buffer. We refer to the encoded media data as "Media Buffer". The Media Buffer is a stream of bytes that may be encoded into a compressed format. 
\item  4 bytes store the CRC (cyclic redundancy code/checksum. The CRC is a network-byte-order CRC-32 computed over the Media Buffer.
\end{enumerate}

\noindent Whenever there are several media data, even of different formats, we store them in a sequential manner each having the aforementioned structure. 

\subsubsection{DDOIF filename extension}
We add the ".ddof" extension to any DDOIF filename. 

\noindent For illustration, we used a set of images of a sport jacket at different views and conditions to populate a single DDOIF file. Via DDOIF API, we stored all these images into as single file in DDOIF format, along with classes defining attributes of the jacket taken from DDOIF dictionary. The file \textit{jacket.ddof} is made publicly available at~\url{https://github.com/morawi/ddoif/tree/master/Examples} and any user can use the DDOIF API to load it. We designed the file format in a way that accepts an unlimited number of media data.

\newcommand{\wsz}{0.16}
\begin{figure}[htp]
  \centering
      \includegraphics[width=\wsz\linewidth]{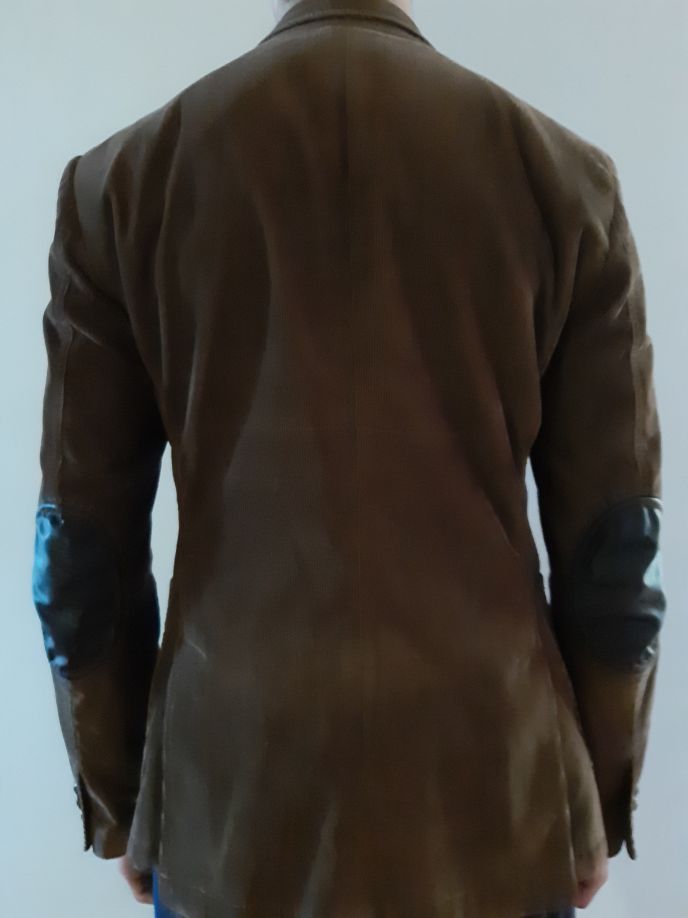}
      \includegraphics[width=\wsz\linewidth]{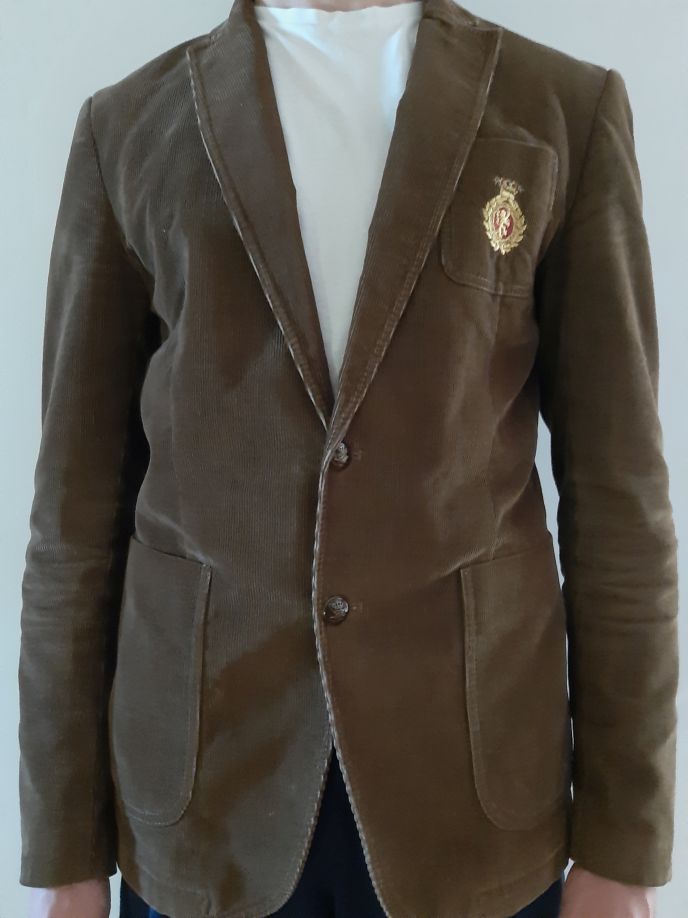}
      \includegraphics[width=\wsz\linewidth]{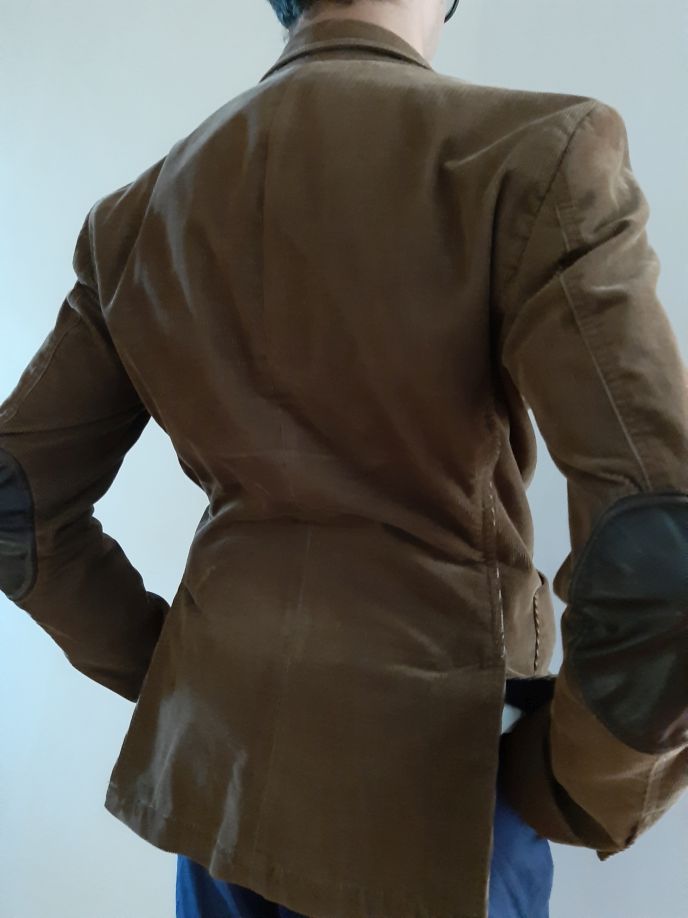}
      \includegraphics[width=\wsz\linewidth]{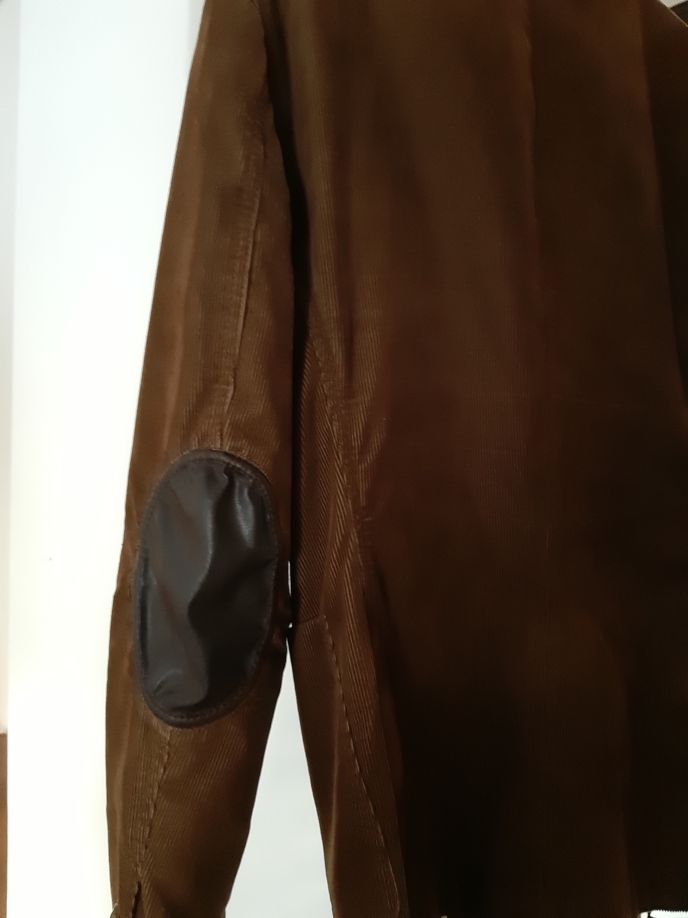}
      \includegraphics[width=\wsz\linewidth]{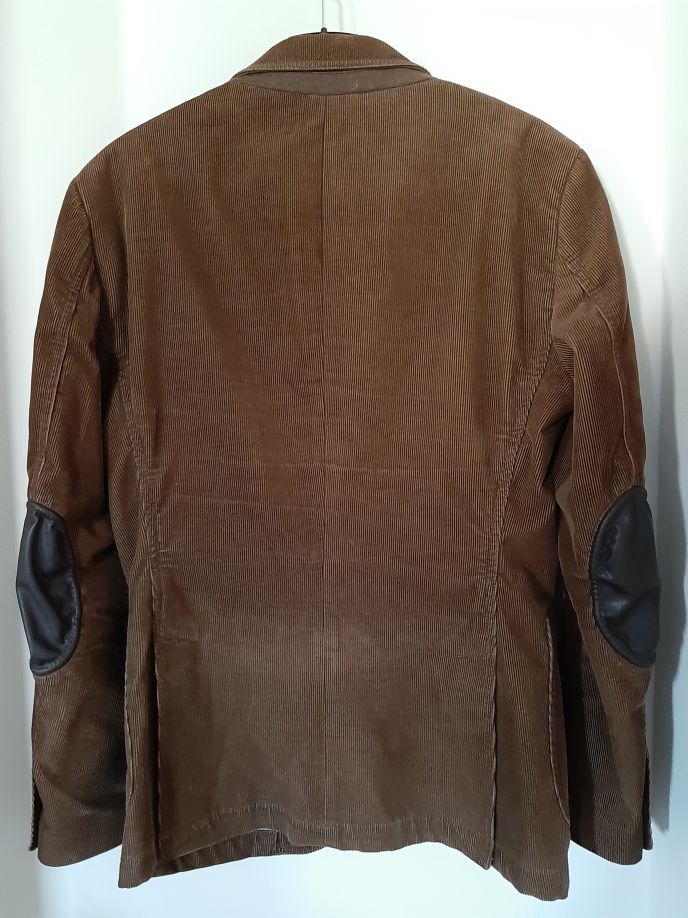}
      \includegraphics[width=\wsz\linewidth]{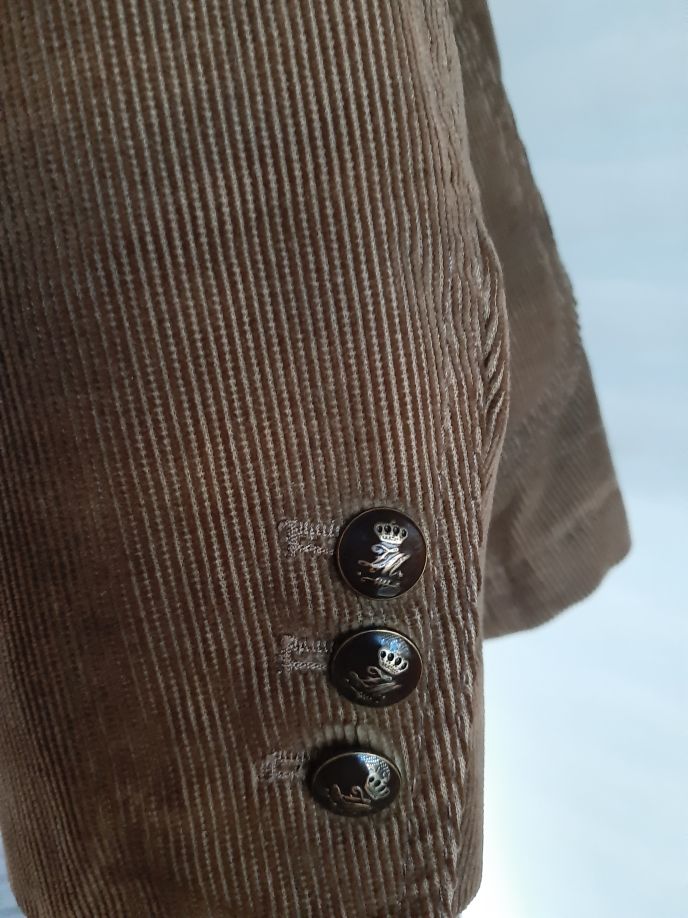}
      \includegraphics[width=\wsz\linewidth]{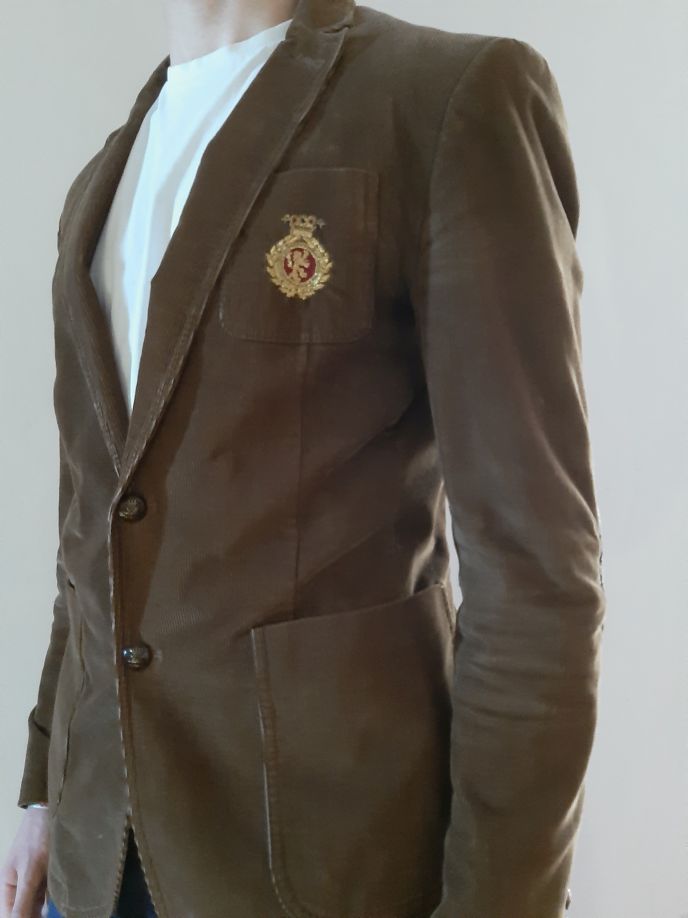}
      \includegraphics[width=\wsz\linewidth]{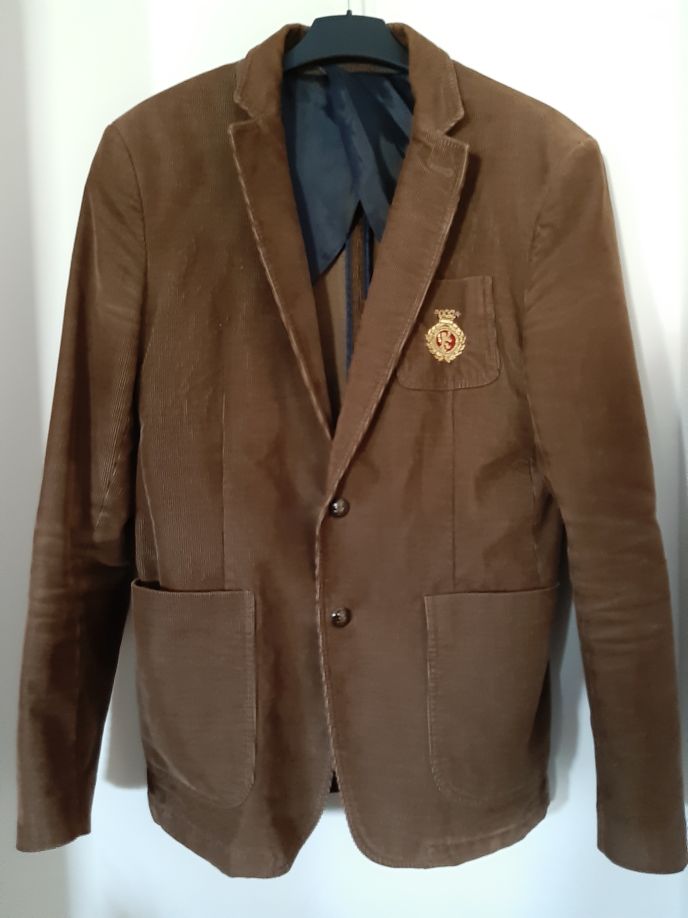}
      \includegraphics[width=\wsz\linewidth]{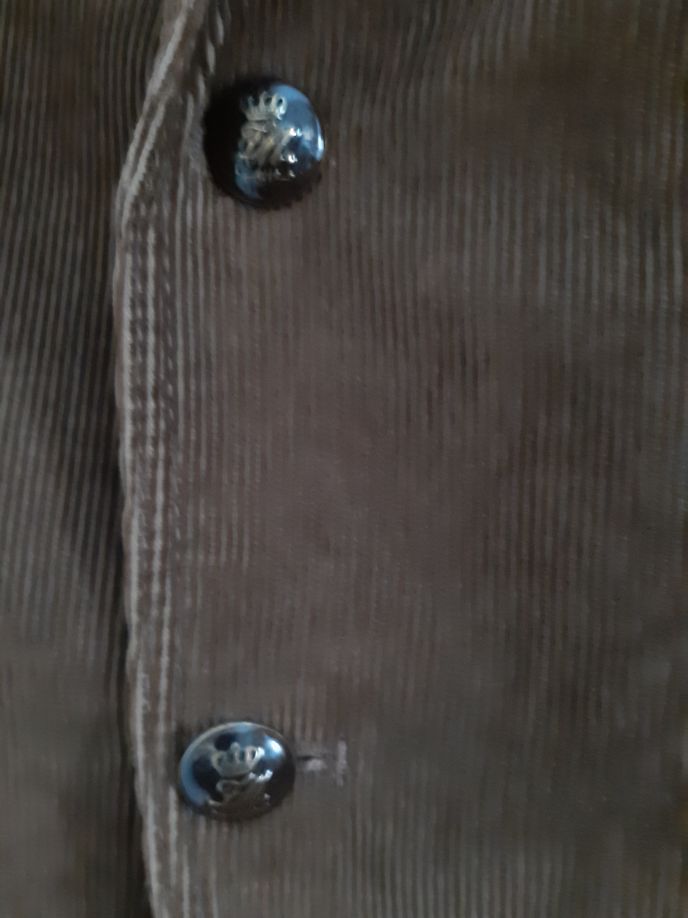}
      \includegraphics[width=\wsz\linewidth]{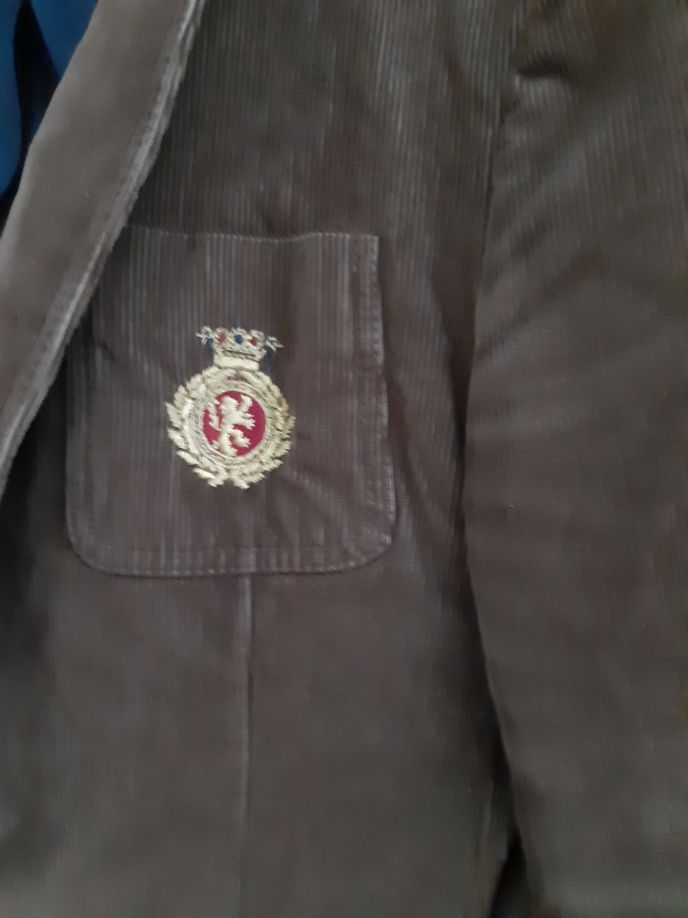}
      \caption{Sport jacket at different views and conditions. All these images have been stored into (a single file) DDOIF format along with classes defining attributes of the jacket taken from DDOIF dictionary. See \textit{jacket.ddof}  in \url{https://github.com/morawi/ddoif/tree/master/Examples}. } 
  \Description{Sport Jacket}
  \label{fig:Jacket-ddoif}
\end{figure}

\section{Use Cases}
In addition to interoperability, DDOIF can be used in several fashion use-cases, including but not limited to: 
\begin{itemize}
    \item Generating datasets with rich annotations. The amount of details, classes and attributes included in a DDOIF file will enable AI companies take advantage and build novel fine-grained classification systems that can be applied to solve different problems. This is because DDOIF not only uses a unified fashion dictionary and highly structured file, but can also have a vintage approach as a bottom-up hierarchy that starts from the early stages of fashion design. As an example, AI companies usually use fashion datasets containing images. There are two cases where fashion data are annotated; the first is for images scrapped from (or provided by) online retailers, such as Amazon ~\cite{Guo2019FashionIQ}; the second is for images scrapped from social media or uploaded by some users as in ~\cite{Zheng2018FashionModaNet}. Although our approach aims at encapsulating fashion data into a single file, it cant still be used to provide accurate annotation. For the social media example the user can (1) grab the DDOIF information from a dedicated mobile / desktop app, or (2) scan the barcode or QR-code attached to the piece of clothing and the DDOIF information can then be imported from a DDOIF centralized data centre. 
    
    \item Improved design, styling and development of fashion items. Fashion design knowledge and items will be available at a large scale. This will help stylists and designers make the best of use of the technology, and dedicated fashion design software will also make use of the technology.
    
    \item Clothing online retrieval based on text and / or images queries. Compared to the simple annotation currently in use, the rich context, based on vast clothing attributes that we are proposing in DDOIF, will enable AI companies build better predicting systems.
    
    \item Fashion data archiving and retrieval. Interoperability is the most important aspect here, and archiving and retrieval will be spread across relevant sectors and other interested parties.
    
    \item Fashion evolution analysis. Storing fashion items in DDOIF format will enable using them in a longitudinal manner if backward compatibility is maintained. The analysis involves repeated observations of the same variables over short or long periods of time. This way, one can investigate how those fashion items have changed over time, which can be helpful for different applications. 
    \item Recommender Systems. This is one of the ultimate applications that can make use of DDOIF format, as the latter will provide comprehensive, interoperable fashion data that are structured according to a standardised protocol.
    
    \item Fashion forensics and brand protection. We propose fashion forensics as a mean to be used for investigating theft of fashion intellectual properties, which to our knowledge first time to be proposed. The aim of these tools is to enhance the role of intellectual property rights in the fashion business. 

\end{itemize}

\section{Discussion and Conclusions}

\subsection{Data interoperability}
The protocol we propose will have a significant impact on Data Management and Software Interoperability used in fashion. It will enhance the capability of different fashion companies to exchange data via a common set of exchange formats, to read and write the same file format, and to use the same protocols. That said, it will help the fashion industry and the IT companies build platforms that have the ability to collect, categorise, store fashion data and provide different tools useful to marketers. Hence, not only that DDOIF aims to provide interoperability within and between the fashion industry and third-party companies, but it can also be used to solve the problem of fashion annotation as well as providing accurate and rich annotations that can never be achieved through crowdsourcing. DDOIF dictionary ensures portability, completeness, efficiency, comparability, compatibility, interchangeability, freedom of legal restrictions, easiness, and integrity. We make DDOIF publicly available in \cite{DDOIFV0.1} (MIT License).
 
\subsection{Impact on the fashion market} 
The protocol will also encourage building a central fashion data management platforms, giving the stakeholders, \eg, marketers, executives, designers, stylists, \etc, the most accurate business and customer information available. Moreover, it will not only provide an opportunity improvement of fashion data collection and exchange, but also interoperability across different platforms and fashion industries that are making use of the AI boom. This, to our knowledge, is the first effort aimed at drafting such a protocol for fashion data. We therefore rely on the relevant scientific and professional community to provide feedback and contribute to improving, modifying, and updating this living document and tool. We are thus keen and open to receive cooperation from the fashion industry and other interested parties to review, enrich and expand DDOIF.
 
\subsection{Fashion data standardisation} 
As we aim to boost fashion data organisation and exchange, DDOIF will also enable non-trivial research activity aligned with the information and knowledge management. The rich textual and media information it provides will make knowledge representation and reasoning from fashion data possible. As a resource, DDOIF will simultaneously provide powerful research for fashion AI and foster completely new areas as well. As a living tool, dictionary and API, we plan to curate and update DDOIF; hopefully, it will one day become an ICT standard. This is a long way to go as it heavily depend on collaboration of many partners from both the fashion industry, AI companies and academia. It is not hard to conjecture that DDOIF will be beneficial for a very long time, as it relates to prominent human needs, clothes and accessories. 

\subsection{How customers will benefit from DDOIF}
What we suggest might be questioned whether users would find this protocol useful in real applications. It is really a challenging task to tag fashion products such as clothes at the retailer end. Even if clothes are tagged with their materials, colour or general styles, these tags do not work well in the application scenarios such as product recommendation -- consumers do not care about these tags (they are more driven by visual contents). DDOIF provides a practical solutions to this problem as the tags are generated at the design phase, and shipped in DDOIF format to the customers along with the clothing item. Tags can be retrieved via bar code that accesses some central database, or a tiny chip shipped with the item. Hence, each customer will possess the clothing item along with all the tags, and she/he can use these tags in application scenarios, as depicted in Fig. \ref{fig:teaser}. Importantly, the protocol is not intended for customers per se, but also for fashion and AI companies that provide enhanced services for customers.  

\subsection{Usefulness of information collected from customers}
For recommender systems, DDOIF tags, alongside the visual content and media of the items, will provide rich information. However, many fashion related businesses have recently been interested in studying how consumers perceive fashion items. They are especially interested in the tags related to subjective feeling (e.g., whether a shoe looks modern, fantastic or outdated). Companies usually collect this information from customers. Such information can also be fed to recommender system learning and update, as these information will be updated to DDOIF via the feedback tags. Retailers have already developed applications to collect such personal preferences, see ~\cite{AmazStch20,StichFix20} for details.

\subsection{Future aspects}
Research user community, which includes industry and academia, is heavily involved in fashion AI. Their interest in DDOIF is likely to be prominent and grow further in the next few years. This is clearly expected as the the fashion industry, which dominates the online market, is a huge business currently worth around US\$2.4 trillion\footnote{These numbers were collected before COVID-19.}. Although DDOIF can help with physical shopping via handset apps, it will make a huge difference to online shoppers as more than 2.14 billion people worldwide are expected to purchase goods and services online.

% https://www.statista.com/statistics/251666/number-of-digital-buyers-worldwide/
% https://www.statista.com/topics/5091/apparel-market-worldwide/

\begin{acks}
This work has received funding from the Elite-S project which is co-funded by ADAPT Centre, Trinity College Dublin and the European Union’s Horizon 2020 programme under the Marie Skłodowska-Curie Grant Agreement No. 801522.
\end{acks}

\bibliographystyle{ACM-Reference-Format}
\bibliography{sample-base}


\begin{thebibliography}{27}

%%% the Web address.

\ifx \showCODEN    \undefined \def \showCODEN     #1{\unskip}     \fi
\ifx \showDOI      \undefined \def \showDOI       #1{#1}\fi
\ifx \showISBNx    \undefined \def \showISBNx     #1{\unskip}     \fi
\ifx \showISBNxiii \undefined \def \showISBNxiii  #1{\unskip}     \fi
\ifx \showISSN     \undefined \def \showISSN      #1{\unskip}     \fi
\ifx \showLCCN     \undefined \def \showLCCN      #1{\unskip}     \fi
\ifx \shownote     \undefined \def \shownote      #1{#1}          \fi
\ifx \showarticletitle \undefined \def \showarticletitle #1{#1}   \fi
\ifx \showURL      \undefined \def \showURL       {\relax}        \fi
% invisible to TeX
\providecommand\bibfield[2]{#2}
\providecommand\bibinfo[2]{#2}
\providecommand\natexlab[1]{#1}
\providecommand\showeprint[2][]{arXiv:#2}

\bibitem[\protect\citeauthoryear{Al-rawi}{Al-rawi}{2020}]%
        {DDOIFV0.1}
\bibfield{author}{\bibinfo{person}{Mohammed Al-rawi}.}
  \bibinfo{year}{2020}\natexlab{}.
\newblock \bibinfo{booktitle}{\emph{ddoif - Digital Data Organization and
  Exchange in Fashion}}.
\newblock ADAPT, Trinity College Dublin.
\newblock
\urldef\tempurl%
\url{https://github.com/morawi/ddoif}
\showURL{%
Retrieved May 26, 2020 from \tempurl}


\bibitem[\protect\citeauthoryear{Amazon}{Amazon}{2020}]%
        {AmazonFashion}
\bibfield{author}{\bibinfo{person}{Amazon}.} \bibinfo{year}{2020}\natexlab{}.
\newblock \bibinfo{booktitle}{\emph{Amazon Fashion}}.
\newblock Amazon.com, Inc.
\newblock
\urldef\tempurl%
\url{https://www.amazon.com/fashion}
\showURL{%
Retrieved May 26, 2020 from \tempurl}


\bibitem[\protect\citeauthoryear{Bel, Liu, Alsheikh, Tang, Pizzi, Henning,
  Singh, Parkhi, and Borisyuk}{Bel et~al\mbox{.}}{2020}]%
        {GrokNet20}
\bibfield{author}{\bibinfo{person}{Sean Bel}, \bibinfo{person}{Yiqun Liu},
  \bibinfo{person}{Sami Alsheikh}, \bibinfo{person}{Yina Tang},
  \bibinfo{person}{Ed Pizzi}, \bibinfo{person}{M. Henning},
  \bibinfo{person}{Karun Singh}, \bibinfo{person}{Omkar Parkhi}, {and}
  \bibinfo{person}{Fedor Borisyuk}.} \bibinfo{year}{2020}\natexlab{}.
\newblock \bibinfo{title}{GrokNet: Unified Computer Vision Model Trunk and
  Embeddings For Commerce}.
\newblock
\newblock
\urldef\tempurl%
\url{https://ai.facebook.com/research/publications/groknet-unified-computer-vision-model-trunk-and-embeddings-for-commerce/}
\showURL{%
Retrieved May 25, 2020 from \tempurl}


\bibitem[\protect\citeauthoryear{Berg, Bell, Paluri, Chtcherbatchenko, Chen,
  Ge, and Yin}{Berg et~al\mbox{.}}{2020}]%
        {FaceBk20}
\bibfield{author}{\bibinfo{person}{Tamara Berg}, \bibinfo{person}{Sean Bell},
  \bibinfo{person}{Manohar Paluri}, \bibinfo{person}{Andrei Chtcherbatchenko},
  \bibinfo{person}{Harry Chen}, \bibinfo{person}{Francis Ge}, {and}
  \bibinfo{person}{Bo Yin}.} \bibinfo{year}{2020}\natexlab{}.
\newblock \bibinfo{title}{Powered by AI: Advancing product understanding and
  building new shopping experiences}.
\newblock
\newblock
\urldef\tempurl%
\url{https://tinyurl.com/yyomybc9}
\showURL{%
Retrieved July 22, 2020 from \tempurl}


\bibitem[\protect\citeauthoryear{Biron}{Biron}{2019}]%
        {AmazStch20}
\bibfield{author}{\bibinfo{person}{Bethany Biron}.}
  \bibinfo{year}{2019}\natexlab{}.
\newblock \bibinfo{title}{Amazon launched a new personal-styling service that
  works a lot like Stitch Fix}.
\newblock
\newblock
\urldef\tempurl%
\url{https://tinyurl.com/yy86ajhj}
\showURL{%
Retrieved July 22, 2020 from \tempurl}


\bibitem[\protect\citeauthoryear{co.}{co.}{2020}]%
        {StichFix20}
\bibfield{author}{\bibinfo{person}{Stich~Fix co.}}
  \bibinfo{year}{2020}\natexlab{}.
\newblock \bibinfo{title}{Personal Styling for Everybody}.
\newblock
\newblock
\urldef\tempurl%
\url{https://www.stitchfix.com/}
\showURL{%
Retrieved August 22, 2020 from \tempurl}


\bibitem[\protect\citeauthoryear{Crowston}{Crowston}{2012}]%
        {Crowston2012AmazonMT}
\bibfield{author}{\bibinfo{person}{Kevin Crowston}.}
  \bibinfo{year}{2012}\natexlab{}.
\newblock \showarticletitle{Amazon Mechanical Turk: A Research Tool for
  Organizations and Information Systems Scholars}. In
  \bibinfo{booktitle}{\emph{Shaping the Future of ICT Research. Methods and
  Approaches}}, \bibfield{editor}{\bibinfo{person}{Anol Bhattacherjee} {and}
  \bibinfo{person}{Brian Fitzgerald}} (Eds.). \bibinfo{publisher}{Springer
  Berlin Heidelberg}, \bibinfo{address}{Berlin, Heidelberg},
  \bibinfo{pages}{210--221}.
\newblock
\showISBNx{978-3-642-35142-6}


\bibitem[\protect\citeauthoryear{E2339}{E2339}{2016}]%
        {AstmDICONDE}
\bibfield{author}{\bibinfo{person}{ASTM E2339}.}
  \bibinfo{year}{2016}\natexlab{}.
\newblock \bibinfo{booktitle}{\emph{Standard Practice for Digital Imaging and
  Communication in Nondestructive Evaluation (DICONDE)}}.
\newblock American Society for Testing and Materials.
\newblock
\urldef\tempurl%
\url{https://www.astm.org/Standards/E2339.htm}
\showURL{%
Retrieved May 10, 2020 from \tempurl}


\bibitem[\protect\citeauthoryear{EasySize}{EasySize}{2020}]%
        {size3}
\bibfield{author}{\bibinfo{person}{EasySize}.} \bibinfo{year}{2020}\natexlab{}.
\newblock
\newblock
\urldef\tempurl%
\url{https://www.easysize.me/}
\showURL{%
Retrieved July 21, 2020 from \tempurl}


\bibitem[\protect\citeauthoryear{Ge, Zhang, Wang, Tang, and Luo}{Ge
  et~al\mbox{.}}{2019}]%
        {Ge2019DeepFashion2AV}
\bibfield{author}{\bibinfo{person}{Yuying Ge}, \bibinfo{person}{Ruimao Zhang},
  \bibinfo{person}{Xiaogang Wang}, \bibinfo{person}{Xiaoou Tang}, {and}
  \bibinfo{person}{Ping Luo}.} \bibinfo{year}{2019}\natexlab{}.
\newblock \showarticletitle{DeepFashion2: A Versatile Benchmark for Detection,
  Pose Estimation, Segmentation and Re-Identification of Clothing Images}. In
  \bibinfo{booktitle}{\emph{The IEEE Conference on Computer Vision and Pattern
  Recognition (CVPR)}}. \bibinfo{publisher}{Computer Vision Foundation / IEEE},
  \bibinfo{address}{Long Beach, California, United States},
  \bibinfo{pages}{5337--5345}.
\newblock


\bibitem[\protect\citeauthoryear{Google}{Google}{2020}]%
        {GoogleArt}
\bibfield{author}{\bibinfo{person}{Google}.} \bibinfo{year}{2020}\natexlab{}.
\newblock \bibinfo{booktitle}{\emph{Googel Art and Culture}}.
\newblock Google LLC.
\newblock
\urldef\tempurl%
\url{https://artsandculture.google.com/project/fashion-making-of}
\showURL{%
Retrieved May 22, 2020 from \tempurl}


\bibitem[\protect\citeauthoryear{Guo, Wu, Gao, Rennie, and Feris}{Guo
  et~al\mbox{.}}{2019}]%
        {Guo2019FashionIQ}
\bibfield{author}{\bibinfo{person}{Xiaoxiao Guo}, \bibinfo{person}{Hui~Qing
  Wu}, \bibinfo{person}{Yupeng Gao}, \bibinfo{person}{Steven~J. Rennie}, {and}
  \bibinfo{person}{Rog{\'e}rio~Schmidt Feris}.}
  \bibinfo{year}{2019}\natexlab{}.
\newblock \showarticletitle{The Fashion IQ Dataset: Retrieving Images by
  Combining Side Information and Relative Natural Language Feedback}.
\newblock \bibinfo{journal}{\emph{ArXiv}}  \bibinfo{volume}{abs/1905.12794}
  (\bibinfo{year}{2019}), \bibinfo{pages}{00--00}.
\newblock


\bibitem[\protect\citeauthoryear{Hou, Wu, Chen, Li, Zheng, and Liu}{Hou
  et~al\mbox{.}}{2019}]%
        {Hou2019RecSys}
\bibfield{author}{\bibinfo{person}{Min Hou}, \bibinfo{person}{Le Wu},
  \bibinfo{person}{Enhong Chen}, \bibinfo{person}{Zhi Li},
  \bibinfo{person}{Vincent~Wenchen Zheng}, {and} \bibinfo{person}{Qi Liu}.}
  \bibinfo{year}{2019}\natexlab{}.
\newblock \showarticletitle{Explainable Fashion Recommendation: A Semantic
  Attribute Region Guided Approach}. In \bibinfo{booktitle}{\emph{IJCAI}}.
  \bibinfo{publisher}{International Joint Conference on Artificial
  Intelligence}, \bibinfo{address}{Macao, China}, \bibinfo{pages}{211--228}.
\newblock


\bibitem[\protect\citeauthoryear{Inc}{Inc}{2018}]%
        {AmazonPatent}
\bibfield{author}{\bibinfo{person}{Amazon~Technologies Inc}.}
  \bibinfo{year}{2018}\natexlab{}.
\newblock \bibinfo{booktitle}{\emph{On demand apparel manufacturing}}.
\newblock Amazon.
\newblock
\urldef\tempurl%
\url{https://patents.google.com/patent/US9623578B1/en}
\showURL{%
Retrieved May 16, 2020 from \tempurl}


\bibitem[\protect\citeauthoryear{ISO}{ISO}{2016}]%
        {iso2016}
\bibfield{author}{\bibinfo{person}{ISO}.} \bibinfo{year}{2016}\natexlab{}.
\newblock \bibinfo{booktitle}{\emph{ISO 18831:2016(en) Clothing — Digital
  fittings — Attributes of virtual garments}}.
\newblock International Organization for Standardization.
\newblock
\urldef\tempurl%
\url{https://www.iso.org/obp/ui/#iso:std:iso:18831:ed-1:v1:en}
\showURL{%
Retrieved May 2, 2020 from \tempurl}


\bibitem[\protect\citeauthoryear{Kahn, Carrino, Flynn, Peck, and Horii}{Kahn
  et~al\mbox{.}}{2007}]%
        {Kahn2007DICOMAR}
\bibfield{author}{\bibinfo{person}{Charles~E. Kahn}, \bibinfo{person}{John~A.
  Carrino}, \bibinfo{person}{Michael~Joseph Flynn}, \bibinfo{person}{Donald~J.
  Peck}, {and} \bibinfo{person}{Steven~C. Horii}.}
  \bibinfo{year}{2007}\natexlab{}.
\newblock \showarticletitle{DICOM and radiology: past, present, and future.}
\newblock \bibinfo{journal}{\emph{Journal of the American College of Radiology
  : JACR}}  \bibinfo{volume}{4 9} (\bibinfo{year}{2007}),
  \bibinfo{pages}{652--7}.
\newblock


\bibitem[\protect\citeauthoryear{Lasserre, Bracher, and Vollgraf}{Lasserre
  et~al\mbox{.}}{2018}]%
        {Studio2ShopFS}
\bibfield{author}{\bibinfo{person}{Julia Lasserre}, \bibinfo{person}{Christian
  Bracher}, {and} \bibinfo{person}{Roland Vollgraf}.}
  \bibinfo{year}{2018}\natexlab{}.
\newblock \showarticletitle{Street2Fashion2Shop: Enabling Visual Search in
  Fashion e-Commerce Using Studio Images}. In \bibinfo{booktitle}{\emph{Pattern
  Recognition Applications and Methods - 7th International Conference, {ICPRAM}
  2018, January 16-18, 2018, Revised Selected Papers}}
  \emph{(\bibinfo{series}{Lecture Notes in Computer Science})},
  \bibfield{editor}{\bibinfo{person}{Maria~De Marsico},
  \bibinfo{person}{Gabriella~Sanniti di~Baja}, {and} \bibinfo{person}{Ana L.~N.
  Fred}} (Eds.), Vol.~\bibinfo{volume}{11351}. \bibinfo{publisher}{Springer},
  \bibinfo{address}{Madeira, Portugal}, \bibinfo{pages}{3--26}.
\newblock
\urldef\tempurl%
\url{https://doi.org/10.1007/978-3-030-05499-1\_1}
\showDOI{\tempurl}


\bibitem[\protect\citeauthoryear{Liang, Lin, Yang, Luo, Huang, and Yan}{Liang
  et~al\mbox{.}}{2016}]%
        {Liang2016FashionCoParse}
\bibfield{author}{\bibinfo{person}{Xiaodan Liang}, \bibinfo{person}{Liang Lin},
  \bibinfo{person}{Wei Yang}, \bibinfo{person}{Ping Luo},
  \bibinfo{person}{Junshi Huang}, {and} \bibinfo{person}{Shuicheng Yan}.}
  \bibinfo{year}{2016}\natexlab{}.
\newblock \showarticletitle{Clothes Co-Parsing Via Joint Image Segmentation and
  Labeling With Application to Clothing Retrieval}.
\newblock \bibinfo{journal}{\emph{IEEE Transactions on Multimedia}}
  \bibinfo{volume}{18} (\bibinfo{year}{2016}), \bibinfo{pages}{1175--1186}.
\newblock


\bibitem[\protect\citeauthoryear{Mr~Jinchun~Yang}{Mr~Jinchun~Yang}{2016}]%
        {isoDigiFitt}
\bibfield{author}{\bibinfo{person}{et.~al. Mr~Jinchun~Yang}.}
  \bibinfo{year}{2016}\natexlab{}.
\newblock \bibinfo{booktitle}{\emph{Clothing sizing systems - size designation,
  size measurement methods and digital fittings}}.
\newblock International Organization for Standardization.
\newblock
\urldef\tempurl%
\url{https://www.iso.org/committee/52374.html}
\showURL{%
Retrieved May 4, 2020 from \tempurl}


\bibitem[\protect\citeauthoryear{NEMA}{NEMA}{2020}]%
        {Dicom}
\bibfield{author}{\bibinfo{person}{NEMA}.} \bibinfo{year}{2020}\natexlab{}.
\newblock \bibinfo{booktitle}{\emph{NEMA PS3 / ISO 12052, Digital Imaging and
  Communications in Medicine (DICOM) Standard}}.
\newblock National Electrical Manufacturers Association, Rosslyn, VA, USA.
\newblock
\urldef\tempurl%
\url{http://www.dicomstandard.org}
\showURL{%
Retrieved May 19, 2020 from \tempurl}


\bibitem[\protect\citeauthoryear{Ollion}{Ollion}{2018}]%
        {HeuriTech}
\bibfield{author}{\bibinfo{person}{Charles Ollion}.}
  \bibinfo{year}{2018}\natexlab{}.
\newblock \bibinfo{booktitle}{\emph{Why computer vision APIs won’t do the
  trick for verticalized applications. Heuritech’s take in Fashion}}.
\newblock Zalando SEs.
\newblock
\urldef\tempurl%
\url{https://bit.ly/3dcCsT6}
\showURL{%
Retrieved May 22, 2020 from \tempurl}


\bibitem[\protect\citeauthoryear{Pennebaker and Mitchell}{Pennebaker and
  Mitchell}{1992}]%
        {Penn92}
\bibfield{author}{\bibinfo{person}{William~B. Pennebaker} {and}
  \bibinfo{person}{Joan~L. Mitchell}.} \bibinfo{year}{1992}\natexlab{}.
\newblock \bibinfo{booktitle}{\emph{JPEG Still Image Data Compression
  Standard}}.
\newblock \bibinfo{publisher}{Van Nostrand Reinhold}, \bibinfo{address}{New
  York}.
\newblock


\bibitem[\protect\citeauthoryear{PopOptiq}{PopOptiq}{2020}]%
        {PopOptiq}
\bibfield{author}{\bibinfo{person}{PopOptiq}.} \bibinfo{year}{2020}\natexlab{}.
\newblock \bibinfo{booktitle}{\emph{PopOptiq Fashion}}.
\newblock PopOptiq.
\newblock
\urldef\tempurl%
\url{https://www.popoptiq.com/fashion/}
\showURL{%
Retrieved May 26, 2020 from \tempurl}


\bibitem[\protect\citeauthoryear{SizerMe}{SizerMe}{2020}]%
        {size2}
\bibfield{author}{\bibinfo{person}{SizerMe}.} \bibinfo{year}{2020}\natexlab{}.
\newblock
\newblock
\urldef\tempurl%
\url{https://sizer.me/}
\showURL{%
Retrieved July 21, 2020 from \tempurl}


\bibitem[\protect\citeauthoryear{Zalando}{Zalando}{2020}]%
        {Zalando}
\bibfield{author}{\bibinfo{person}{Zalando}.} \bibinfo{year}{2020}\natexlab{}.
\newblock \bibinfo{booktitle}{\emph{Zalando SEs}}.
\newblock Zalando SEs.
\newblock
\urldef\tempurl%
\url{https://www.zalando.com/}
\showURL{%
Retrieved May 22, 2020 from \tempurl}


\bibitem[\protect\citeauthoryear{ZeeKit}{ZeeKit}{2020}]%
        {size1}
\bibfield{author}{\bibinfo{person}{ZeeKit}.} \bibinfo{year}{2020}\natexlab{}.
\newblock
\newblock
\urldef\tempurl%
\url{https://zeekit.me/}
\showURL{%
Retrieved July 23, 2020 from \tempurl}


\bibitem[\protect\citeauthoryear{Zheng, Yang, Kiapour, and Piramuthu}{Zheng
  et~al\mbox{.}}{2018}]%
        {Zheng2018FashionModaNet}
\bibfield{author}{\bibinfo{person}{Shuai Zheng}, \bibinfo{person}{Fan Yang},
  \bibinfo{person}{M.~Hadi Kiapour}, {and} \bibinfo{person}{Robinson
  Piramuthu}.} \bibinfo{year}{2018}\natexlab{}.
\newblock \showarticletitle{ModaNet: A Large-Scale Street Fashion Dataset with
  Polygon Annotations}. In \bibinfo{booktitle}{\emph{Proceedings of the 26th
  ACM International Conference on Multimedia}} (Seoul, Republic of Korea)
  \emph{(\bibinfo{series}{MM ’18})}. \bibinfo{publisher}{Association for
  Computing Machinery}, \bibinfo{address}{New York, NY, USA},
  \bibinfo{pages}{1670–1678}.
\newblock
\showISBNx{9781450356657}
\urldef\tempurl%
\url{https://doi.org/10.1145/3240508.3240652}
\showDOI{\tempurl}


\end{thebibliography}

\end{document}